\documentclass[
  a4paper, 
  10pt, twocolumn,
  draft
]{article}

\title {\LARGE\bf 
                Statistics of delta peaks in the spectral density \\
                of large random trees
}
\author { 	O. Golinelli
\bigskip
\\ \ad		Physique Th\'eorique, Cea Saclay, 91191 Gif, France
\\ \ad		email: golinelli@cea.fr
}

\date{\normalsize	January 23, 2003
\\			Preprint T03/008 ; arXiv:cond-mat/0301437}

\newcommand  {\ad}{\normalsize\em}	
\usepackage[final]{graphicx}
\newcommand{\figwidth}{\columnwidth} 

\begin{document}
\twocolumn[\hsize\textwidth\columnwidth\hsize\csname @twocolumnfalse\endcsname
\maketitle

\begin{abstract} \normalsize
We present an analysis of the spectral density of the adjacency matrix of
large random trees.  We show that there is an infinity of delta peaks at
all real numbers which are eigenvalues of finite trees.  By exact
enumerations and Monte-Carlo simulations, we have numerical estimations of
the heights of peaks.  In the large tree limit, the sum of their heights is
$0.19173 \pm 0.00005$.  Moreover all associated eigenvectors are strictly
localized on a finite number of nodes.  The rest of the spectral density is
a function which vanishes at all positions of peaks, which are a dense
subset of real numbers: so this function is almost everywhere
discontinuous.


\end{abstract}
\vskip2pc ]              

\pagebreak
\section{Introduction}

Many models of graphs have been investigated in the last decades.  This
interest is mainly motivated by their numerous areas of applications in
physics (propagation and percolation in disordered media, quantum gravity,
etc.)  and in other branches of science: combinatorial optimization,
queuing theory, computers networks, interactions between biological
molecules, quantum chemistry and many others.  But their study by methods
and concepts of statistical mechanics is more recent.  The point of view of
statistical physicists is generally different from the one of
mathematicians, in particular by studying averaged quantities in the
``thermodynamic'' limit of infinite graphs rather than specific quantities
to a given finite graph.  Moreover some results have been obtained by
Monte-Carlo methods, which are often regarded as ``heretical'' by
mathematicians who prefer exact and proved results.

We can describe a graph as a set of $n$ nodes (or vertices) with
interactions between pairs of nodes, represented by a $n \times n$ matrix.
Then we are naturally interested by the spectrum of this adjacency matrix.
See for example the book [Cvetkovic and al., 1995] for the theory and
applications of these spectra.  Many works have been done about the
singularities of the spectral density and their connections with the
localization of eigenvectors, also called quantum percolation [Evangelou,
1983; Mirlin, 2000; Bauer and Golinelli, 2001; id. 2001a].  The spectrum of
the Laplacian matrix has also been studied [Biroli and Monasson, 1999].

In particular, it is known [Kirkpatrick and Eggarter, 1972] that {\em
strictly} localized eigenvectors (i.e. with a finite number of non-zero
coordinates) can appear on finite parts of large connected graph.  Moreover
the associated eigenvalues contribute to delta peaks in the spectral
density.  But the strict localization can occur only on parts of graph with
special patterns.  The motivation of our work is to study quantitatively
this phenomenon for large (i.e in the infinite size limit) connected
graphs.

As we have no reasons to choose a particular graph, we turn to models of
random graphs.  Of course, quantitative results depend on the particular
choice of the model.  Our first idea is to study the famous Erd\"os-R\'enyi
model of random graph [Erd\"os and R\'enyi, 1960] but it presents a
default: a graph on $n$ nodes consists (in the high connectivity phase) of
a giant connected component with $O(n)$ nodes on which strict localization
can occur, plus a number $O(n)$ of finite connected components on which
eigenvectors are always strictly localized.  As we are not interested by
these ones, we must consider either a modified Erd\"os-R\'enyi model by
keeping only the giant component, or an other model.

We prefer the random labeled tree model described in Sect.~\ref{sec:def},
motivated by the following considerations.  Firstly large Erd\"os-R\'enyi
graphs on $n$ nodes have circuits (or loops) with length $O(\ln n)$, so
they have locally a tree shape.  As the strict localization occurs
statistically on small number of nodes, it is little sensitive to circuits.
Furthermore by definition a tree is always connected and we dismiss the
problems due to small components.  And finally a random tree is simpler to
generate than other models of connected graphs.  In this work, the
interaction matrix is the adjacency matrix of the tree: it is the
Hamiltonian of a particle which hops at each time step from one node to a
connected node.  Remark that it is different to the Laplacian matrix which
describes the diffusion of a particle with continuous time.

As analytical methods give only partial results, we use two numerical
methods: exact enumeration of small trees, described in Sect.~\ref{sec:enu}
and Monte-Carlo simulations for large trees, described in
Sect.~\ref{sec:pmc}.  Qualitative description of the spectral density is
done in Sect.~\ref{sec:sd}; existence of an infinity of delta peaks is
proved in Sect.~\ref{sec:existence} and their statistics is given in
Sect.~\ref{sec:stat}. Appendix~\ref{sec:symspec} contains details about the
use of symmetry for the numerical computation of the spectra.  In
Appendix~\ref{sec:density}, we show that the set of eigenvalues of finite
trees are a dense subset of the real numbers.  Finally
Appendix~\ref{sec:pure} explains the difficulties encountered to extend
analytical results obtained in a previous work [Bauer and Golinelli, 2000].

\section{Definitions and generalities}

\label{sec:def}

In this article, unless otherwise stated, the term {\em tree} refers to a
{\em labeled tree}.  A {\em (labeled) tree} on $n$ {\em nodes} is a
connected graph with nodes (or vertex) labeled $\{1,2,\dots,n\}$ linked by
$n-1$ simple (i.e. undirected, loopless and not multiple) {\em edges}.
Consequently a tree is without circuit (or polygon).  Two nodes are called
{\em adjacent} or {\em neighboring} if they are connected by an edge.  A
{\em leaf } is a node with only one neighbor.  Note that a tree is a
bipartite graph: the set of nodes can be partitioned into two subsets so
that adjacent nodes are in different subsets.

A theorem due to Cayley says that the number of different labeled trees on
$n$ nodes is
\begin{equation}
  T(n) = n^{n-2}.
\end{equation}
A simple proof uses the Pr\"ufer coding explained in
Sect.~\ref{sec:pmc}. See the book [Van Lint and Wilson, 1992] for a general
presentation.  

The {\em adjacency matrix} of any tree on $n$ nodes is the $n \times n$
square matrix $A$ such that $A_{i,j} = 1$ if nodes $i$ and $j$ are adjacent
and 0 otherwise.  Then $A$ is symmetric, with zeroes on the diagonal.  The
walks on a tree are counted by its adjacency matrix: the number of walks of
length $k$ starting at node $i$ and finishing at node $j$ is $(A^k)_{i,j}$.
Similarly, $\mbox{Tr} A^k$ is the number of closed walks of length $k$.

The spectrum (set of eigenvalues) of the adjacency matrix $A$ of a tree $T$
is more simply called the {\em spectrum of} $T$.  By definition, $\lambda$
is an eigenvalue associated with the eigenvector $V = (V_1, V_2, \dots,
V_n)$ if $A V = \lambda V$.  So the eigenvalue equation on the
node $i$ is
\begin{equation}
  \lambda V_i  = \sum_{j \mbox{ \scriptsize adj } i} V_j,
  \label{eq:lambda}
\end{equation}
where the sum runs over the nodes $j$ adjacent to $i$.  As $A$ is
symmetric, its spectrum consists of real eigenvalues with a complete
orthogonal basis of real eigenvectors.  Furthermore, the spectrum is
symmetric with respect to zero because a tree is bipartite: each eigenmode
$(\lambda, V)$ has a partner $(-\lambda, V')$, where the vectors $V$ and
$V'$ are equal on the nodes of one subset and opposite on the nodes of the
other subset.  See Appendix~\ref{sec:symspec} for an application of this
property to the numerical computation of the spectrum.

The {\em symmetry factor} of any tree is the number of permutations of the
nodes that leave invariant this tree.  Two trees are {\em isomorphic} if
they differ only by a permutation of their nodes, or equivalently if their
adjacency matrices differ only by a permutation of rows and columns.  The
number of trees isomorphic to any tree on $n$ nodes with symmetry factor
$S$ is $n!/S$.  The term {\em labeled} emphasizes that we are not
identifying isomorphic trees.  Therefore an {\em unlabeled tree} is an
isomorphism class of labeled trees.  

Clearly, isomorphic trees are cospectral (i.e. they have the same
spectrum), but the reciprocal is wrong: the proportion of trees on $n$
nodes which have a cospectral but non isomorphic partner goes to 1 when $n$
becomes large [Cvetkovic and al., 1995, ch. 6].

\begin{table*}  \centering
  \begin{tabular}{l|l|c|r|l}
    $n$ & tree & & $n!/S$ & spectrum \\
    \hline 
      1 & $T_1$ &
      \begin{picture}(10,12)(-5,-3) \put (0,0){\circle*{5}} \end{picture}
      & 1 & \{0\} \\
    \hline
      2 & $T_2$ & 
      \begin{picture}(25,12)(-5,-3) 
         \put (0,0) {\circle*{5}} 
         \put (0,0) {\line(1,0){15}}
         \put (15,0) {\circle*{5}} 
      \end{picture}
      & 1 & \{-1,\ 1\} \\
    \hline
      3 & $T_3$ &       
      \begin{picture}(40,12)(-5,-3) 
         \put (0,0) {\circle*{5}} 
         \put (0,0) {\line(1,0){15}}
         \put (15,0) {\circle*{5}} 
         \put (15,0) {\line(1,0){15}}
         \put (30,0) {\circle*{5}} 
      \end{picture}
      & 3 & $\{-\sqrt{2},\ 0,\ \sqrt{2} \}$ \\
    \hline
      4 & $T_{4,1}$ &        
      \begin{picture}(36,23)(-5,-3) 
         \put (0,0) {\circle*{5}} 
         \put (0,0) {\line(1,0){15}}
         \put (15,0) {\circle*{5}} 
         \put (15,0) {\line(1,1){11}}
         \put (26,11) {\circle*{5}} 
         \put (15,0) {\line(1,-1){11}}
         \put (26,-11) {\circle*{5}} 
      \end{picture}
      & 4 & $\{ -\sqrt{3},\ 0,\ 0,\ \sqrt{3} \}$ \\
      & $T_{4,2}$ &         
      \begin{picture}(55,20)(-5,-3) 
         \put (0,0) {\circle*{5}} 
         \put (0,0) {\line(1,0){15}}
         \put (15,0) {\circle*{5}} 
         \put (15,0) {\line(1,0){15}}
         \put (30,0) {\circle*{5}} 
         \put (30,0) {\line(1,0){15}}
         \put (45,0) {\circle*{5}} 
      \end{picture}
      & 12 & $\{ \pm (\sqrt{5}-1)/2,\ \pm (\sqrt{5}+1)/2 \}$ \\
    \hline
      5 & $T_{5,1}$ &        
      \begin{picture}(40,27)(-5,-3) 
         \put (0,0) {\circle*{5}} 
         \put (0,0) {\line(1,0){15}}
         \put (15,0) {\circle*{5}} 
         \put (15,0) {\line(1,0){15}}
         \put (30,0) {\circle*{5}} 
         \put (15,0) {\line(0,1){15}}
         \put (15,15) {\circle*{5}} 
         \put (15,0) {\line(0,-1){15}}
         \put (15,-15) {\circle*{5}} 
      \end{picture}
      & 5 & $\{ -2,\ 0,\ 0,\ 0,\ 2 \}$ \\
      & $T_{5,2}$ &        
      \begin{picture}(51,23)(-20,-3) 
         \put (-15,0) {\circle*{5}} 
         \put (-15,0) {\line(1,0){15}}
         \put (0,0) {\circle*{5}} 
         \put (0,0) {\line(1,0){15}}
         \put (15,0) {\circle*{5}} 
         \put (15,0) {\line(1,1){11}}
         \put (26,11) {\circle*{5}} 
         \put (15,0) {\line(1,-1){11}}
         \put (26,-11) {\circle*{5}} 
      \end{picture}
      & 60 & $\{0,\ \pm \sqrt{2-\sqrt{2}},\ \pm \sqrt{2+\sqrt{2}} \}$ \\
      & $T_{5,3}$ &         
      \begin{picture}(70,20)(-20,-3) 
         \put (-15,0) {\circle*{5}} 
         \put (-15,0) {\line(1,0){15}}
         \put (0,0) {\circle*{5}} 
         \put (0,0) {\line(1,0){15}}
         \put (15,0) {\circle*{5}} 
         \put (15,0) {\line(1,0){15}}
         \put (30,0) {\circle*{5}} 
         \put (30,0) {\line(1,0){15}}
         \put (45,0) {\circle*{5}} 
      \end{picture}
      & 60 & $\{ -\sqrt{3},\ -1,\ 0,\ 1,\ \sqrt{3} \}$ \\
  \end{tabular}
  \caption{\em
    Spectra of trees on $n \le 5$ nodes.  The column $n!/S$ gives the
    number of labeled trees isomorphic to the drawing. 
  }
  \label{tab:spectra}
\end{table*}

The trees on $n \le 5$ nodes (and their spectra) are listed in
Table~\ref{tab:spectra}.  A larger table for $n \le 10$ is in [Cvetkovic
and al., 1995].

In order to speak about average properties of trees (for example their
spectral density), we turn the set of $T(n)$ trees on $n$ nodes into
probability space with equiprobable elements, called {\em random (labeled)
trees}.  In other words, a random tree on $n$ nodes is randomly chosen
among the $T(n)$ trees.

Note that the models of random {\em labeled} trees and of random {\em
unlabeled} trees are not equivalent: trees with a high symmetry factor are
less probable with the labeled model.  For example, we have numerically
computed that the average fraction of the spectrum occupied by the
eigenvalue 0 in tree on $n=23$ nodes is 0.1437\dots\ for a labeled tree and
0.2329\dots\ for an unlabeled tree.  By considering the convergence, the
size $n=23$ is sufficiently large to so that a gap exists in the large
$n$ limit.  It can be explained because most of eigenvectors with
$\lambda=0$ have only two non-zero coordinates which are opposite and
localized on two leaves adjacent to a same node.  As this pattern gives a
symmetry factor 2, eigenvalues zero are less frequent in the labeled model.

\section{Numerical methods}

\label{sec:nm}

We use two numerical methods: full enumeration for small trees and
Monte-Carlo simulations for large trees.

\subsection{Enumeration}

\label{sec:enu}

If we want to enumerate all the $T(n)=n^{n-2}$ trees on $n$ nodes with a
computer (for example with the Pr\"ufer coding), the problem becomes very
hard as soon as $n \approx 10$.  It is faster to enumerate all the
unlabeled (i.e non-isomorphic) trees.  For each unlabeled tree $T$ on $n$
nodes, the spectrum (independent of labels on the nodes) is numerically
computed.  As $T$ represents a class of isomorphic labeled trees, the
spectrum is counted with a multiplicity equal to the number of ways of
labeling $T$. i.e. $n!$ divided by its symmetry factor.

Let us call $t(n)$ the number of unlabeled trees on $n$ nodes.  The
sequence $t(1), t(2), \dots,$ starts with 1, 1, 1, 2, 3, 6, 11, 23,
47, 106, \dots  Since works of Jordan, Cayley, Polya and Otter, we known
[Knuth, 1997, p. 388] that $t(n)$ grows exponentially like
\begin{equation}
  t(n) \sim \beta\ \frac{\alpha^n}{n^{5/2}},
\end{equation}
with $\alpha = 2.955765285652\dots$ and $\beta = 0.5349496061\dots$ 

To enumerate the $t(n)$ unlabeled trees, we used a simplified version of
the WROM algorithm [Wright et al., 1986] with computation time ${\cal
O}(n^2 t(n))$, and not ${\cal O}(t(n))$ as the original version.  As the
computation time for the spectrum of one $n \times n$ matrix is ${\cal
O}(n^3)$, and ${\cal O}(n^2)$ or less for the symmetry factor in the worst
case, we see that the slow part is not the enumeration work, but the
computation of the $t(n)$ spectra.  For the same reason, we did not find
useful to work with algorithms like the recursive one of [Li
and Ruskey, 1999]

Thus, in this work, we have enumerated trees up to $n=23$, where $t(23) =
14\;828\;074$.  As the computational time grows exponentially, it is
difficult to enumerate spectra of trees with much more nodes.

\subsection{Pr\"ufer Monte-Carlo method}

\label{sec:pmc}

To study large trees, the full enumeration is impossible.  The
Monte-Carlo method consists by randomly generating a set of $S$ independent
trees, then measuring some properties and averaging them.  Of course, the
results have random noise, which can be estimated with usual formulae of
statistics.  In good cases, the noise decreases as $1/\sqrt{S}$.

Main points are to generate trees with the right probability law and with
an efficient algorithm.  For example, let us consider the naive method which
generates a random graph by choosing $n-1$ edges between $n$ nodes, then
rejects this graph if not a tree.  As trees are very rare among the set of
graphs, this method is not efficient.  

If we modify the procedure by rejecting edges which close a cycle, we
generate at each time a tree but now with a non uniform probability law.
For example, for trees on 4 nodes, Prob$(T_{4,1})=1/15$ and
Prob$(T_{4,2})=11/180$ instead of $1/16$.  For $n=5$,
Prob$(T_{5,1})=1/105$, Prob$(T_{5,2}) = 127/15120 \approx 1/119$ and
Prob$(T_{5,3}) = 113 / 15120 \approx 1/134$ instead of $1/125$.  We see
that this procedure has a bias which favors trees with highly connected
nodes.  More seriously the bias grows exponentially with the size: the
``star'' tree (made of one central node connected to $n-1$ peripheral
nodes) has a probability $1/(2n-3)!!$ instead of $1/n^{n-2}$.

So we use the Pr\"ufer Monte-Carlo method, which is efficient and not
biased.  We first describe the Pr\"ufer coding, but without giving a proof.
Let $T$ be a labeled tree on $n$ nodes.  The Pr\"ufer coding consists in
removing successively one leaf at each step.  We start with $T_1=T$.  For
$1\le i \le n-1$, the step $i$ is the following: let $b_i$ be the leaf of
$T_i$ with the smallest label, let $a_i$ be the neighbor of $b_i$ and let
$T_{i+1}$ be the tree obtained by deleting from $T_i$ the leaf $b_i$ and
the edge $\{b_i,a_i\}$.  The Pr\"ufer code of $T$ is the sequence $(a_1,
a_2, \dots, a_{n-2})$.  As $a_{n-1}=n$ necessarily, it is not included in
the code.

Remark that the sequence $(b_1,b_2,\dots,b_{n-1})$ is always a permutation
of $(1,2,\dots,$ $n-1)$.  In contrast, repetitions can occur among the
code.  More precisely, the number of neighbors of the node $j$ is the
number of $j$ in the code, plus one.  Consequently the leaves are the nodes
which never appear in the code.

To reverse this procedure, start with an arbitrary code $(a_1, a_2, \dots,
a_{n-2})$ with $1 \le a_i \le n$.  By convenience, we complete with
$a_{n-1} = n$.  We built a tree $T$ with the following iterative procedure:
for $1\le i \le n-1$, at step $i$, let $b_i$ be the least number in $[1,n]$
not in $(b_1,b_2,\dots,b_{i-1})$ and not in $(a_i,\dots, a_{n-1})$.  The
tree $T$ is made of the $n-1$ edges $\{b_i,a_i\}$.

Then the Pr\"ufer coding is an one-to-one correspondence between the labeled
trees on $n$ nodes and the $n^{n-2}$ codes $(a_1, a_2, \dots, a_{n-2})$
with $1 \le a_i \le n$.  This proves in particular that $T(n) = n^{n-2}$.

This coding gives an easy Monte-Carlo method: a random labeled tree on $n$
nodes is built by choosing $n-2$ independent random integer numbers in
$[1,n]$ and by applying the reverse procedure.  Then the spectrum is
computed.  As usual with Monte-Carlo methods, the averages are done with as
many independent trees as possible, in order to reduce the statistical
noise.  Of course the eigenvalues are not independent: for example, each
$\lambda \ne 0$ is generated simultaneously with $-\lambda$.  So the
estimators of variance must be calculated by considering that the
independent events are the generated trees, then their whole spectra, but
not the eigenvalues separately.

As the computation time to generate a random code and then the
corresponding tree is ${\cal O}(n)$, it is possible to obtain trees on
several millions nodes.  But the slow part is always the computation of the
spectrum which needs ${\cal O}(n^3)$ per tree (see details in
Appendix~\ref{sec:symspec}).  This limits to several thousands nodes.

We simulate different sizes to evaluate finite size effects.  For each size
$n =$ 30, 50, 100, 200, 500, 1000, 2000, 5000, 10000, we obtained
$m=30.000.000$ eigenvalues, i.e we simulated $m/n$ random labeled trees of
size $n$.  The main cpu time consumption was for the larger size.  We
judged that it is not interesting to simulate larger trees (with the same
global cpu time): $m$ would be smaller and the fluctuations bigger.

In principal we can simulate {\em random unlabeled trees} with the Pr\"ufer
Monte-Carlo method.  This procedure generates trees with uniform
probability if they are labeled, but non uniform if they are considered as
unlabeled.  So we can correct this bias by giving a weight on each
generated tree equal to its symmetry factor, inversely proportional to its
probability.  We have not simulated unlabeled trees, but it is probable
that new problems appear when $n$ is large: as very few trees with very
large weights dominate the averages, the fluctuations are large and
non-gaussian.  This is similar to the simulation of the low temperature
phase of an Ising model by generating random independent up or down spins,
i.e. at infinite temperature.

\section{Spectral density of large random trees}

\label{sec:sd}

By definition, the {\em spectral density} (or {\em density of states}) of
any tree $T$ on $n$ nodes with spectrum $\{ \lambda_1, \lambda_2, \dots
\lambda_n \}$ is the distribution
\begin{equation}
  \rho_T(x) = \frac{1}{n} \sum_{i=1}^n \delta(x - \lambda_i)
\end{equation}
and the average spectral density of random trees on $n$ nodes is
\begin{equation}
  \rho_n(x) = \frac{1}{n^{n-2}} \sum_{T_n} \rho_{T_n}(x)
\end{equation}
where the sum runs over the $n^{n-2}$ trees on $n$ nodes.  The bounds of
the support of $\rho_n(x)$ are $\max\{\lambda_i\} = \sqrt{n-1}$ and
$\min\{\lambda_i\} = - \sqrt{n-1}$, which are eigenvalues of the ``star''
tree, made of one central node and $n-1$ peripheral nodes.  We are mainly
interested by the asymptotic distribution $\rho(x)$ when $n$ is large, in
the spirit of the statistical physics.

\subsection{Moments of the spectral density}

We consider $\mu_{k,n} = \int dx \ x^k \ \rho_n(x)$, the $k$'th moment of
the average spectral density $\rho_n$.  It is related to the number of
closed walks of length $k$ on the trees on $n$ nodes by
\begin{equation}
  n^{n-1} \mu_{k,n} =  \sum_{T_n} \mbox{Tr}\ T_n^k
\end{equation}
where the symbol $T_n$ represents at the same time a tree on $n$ nodes and
its adjacency matrix. 

As trees are bipartite, their spectra are symmetric with respect to zero
and the odd moments $\mu_{2k+1,n} = 0$.  Furthermore, the first even
moments $\mu_{2k,n}$ can be calculated by enumeration of walks.  For $k=2$,
a closed walk of length 2 is just a ``round trip'' on an edge: as a tree
has $n-1$ edges, $\mbox{Tr}\ T_n^2 = 2(n-1)$ for each tree, then $\mu_{2,n}
= 2 - 2/n$.  The edges visited by a closed walk of length $2k$ on a tree
form a subtree with $l\le k$ edges.  By erasing the $l$ visited edges, the
rest (i.e the spectator edges) splits up into a ``rooted forest'' with $n$
nodes and $m=l+1$ rooted trees. The number of such rooted forests is
\begin{equation}
  F_{m,n} = n^{n-m}\ \frac{(n-1)!}{(n-m)! (m-1)!}.
\end{equation}
The number of closed walks of length $2k$ visiting all nodes of trees with
$l$ edges is $(l+1)!\ {\cal J}_{k,l}$ where the integer numbers ${\cal
J}_{k,l}$ can be computed with a recursion relation: see Eq.~8 and Table~1
of [Bauer and Golinelli, 2001a, p 307].  Consequently, [Bauer, 2000]
\begin{equation}
  \mu_{2k,n} = \sum_{l=1}^k  (l+1)\ {\cal J}_{k,l} \ 
                 \frac{(n-1)(n-2)\dots(n-l)}{n^l},
\end{equation}
valid for any $n \ge 1$ and the first ones are
\begin{eqnarray}
  \mu_{2,n}  &=& 2 - \frac{2}{n} , \\
  \mu_{4,n}  &=& 8 - \frac{20}{n} + \frac{12}{n^2} , \\
  \mu_{6,n}  &=& 40 - \frac{176}{n} + \frac{256}{n^2} - \frac{120}{n^3}.
\end{eqnarray}
The large-$n$ limits of the $\mu_{k,n}$ are
\begin{equation}
  \mu_{2k} = \sum_{l=1}^k (l+1)\ {\cal J}_{k,l}\ \ \mbox{ and }\ \ 
  \mu_{2k+1} = 0.
\end{equation}
The $\mu_{k}$ are the moments of the asymptotic distribution $\rho(x)$ of
the spectral density for large random trees.  Note that $\rho(x)$ is
defined without the need to scale it with a power of $n$.  It is an
``universal'' distribution in the sense that it depends of no parameter,
but only on the definition of the labeled trees.

The sequence $(\mu_{2k})_{k\ge 1}$ starts [Bauer, 2000] with 2, 8, 40, 226,
1384, 8992, 61212, 433136, etc. and we are not aware of any other problem
involving the same sequence.  It can be proved [Bauer and Golinelli, 2001a]
that the growth of the coefficients ${\cal J}_{k,l}$, and then of the
$\mu_{2k}$, is sufficiently slow to characterize $\rho(x)$, and
sufficiently fast to assure that the support of $\rho(x)$ is unbounded.

\subsection{Delta peaks and continuous distribution}

\begin{figure}
  \centering
  \includegraphics [width=\figwidth] {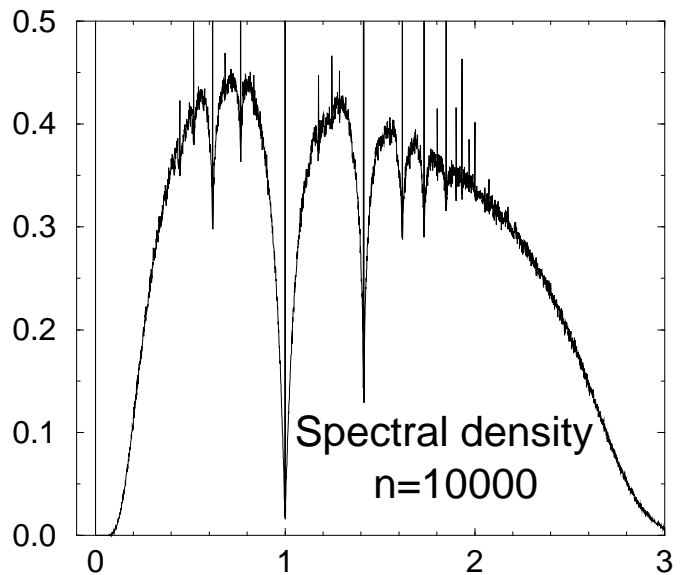}
  \caption{\em 
    Average spectral density $\rho_n(x)$ of labeled trees on $n=10000$
    nodes as function of $x$, represented by a histogram of Monte-Carlo
    eigenvalues.  The bin width is 0.001.  Because of its symmetry, the
    histogram is folded up around zero.
  }
  \label{fig:rho}
\end{figure}

It is not easy to extract accurate local information on the asymptotic
spectral density $\rho(x)$ from the knowledge of a finite number of
moments.  So we have recourse to numerical computations as described in
Sect.~\ref{sec:nm}.  A histogram of Monte-Carlo eigenvalues for $n=10000$
is shown on Fig.~\ref{fig:rho}.  As the spectra of trees are always
symmetric with respect to zero, the histogram is folded up around zero for
convenience sake.

\begin{figure}
  \centering
  \includegraphics [width=\figwidth] {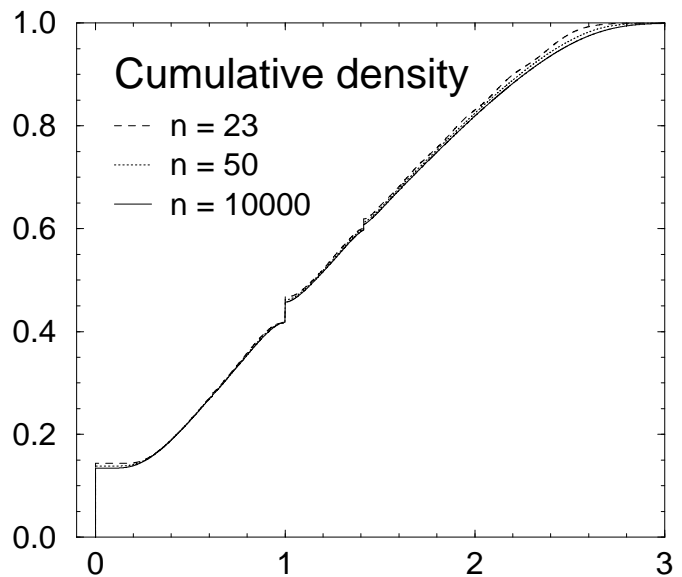}
  \caption{\em 
    Cumulative spectral density of labeled trees on $n=23$, $50$ and
    $10000$.  The convergence is as $1/n$.  The vertical steps correspond
    with delta peaks.  Because of its symmetry, the density is folded up
    around zero.
  }
  \label{fig:cumul}
\end{figure}

Only one choice of size $n$ is represented on Fig.~\ref{fig:rho} because
the dependence on $n$ is small.  In order to give an idea of these finite
size effects, we show on Fig.~\ref{fig:cumul} the (folded) {\em cumulative}
density $R_n(x) = \int_{-x}^x \rho_n(y)\ dy$ for several choices on $n$.
We have observed that $R_n(x)$ converges as $1/n$; for example, the values
of the maximal distance $D_n = \max_x |R_n(x) - R_{10000}(x)|$ are: $D_{23}
\approx 0.021$, $D_{50} \approx 0.0077$, $D_{100} \approx 0.0037$ and
$D_{1000} \approx 0.00042$.  So we can consider that $n=10000$ is
sufficiently large to observe on Fig.~\ref{fig:rho} a distribution very
close to the asymptotic distribution $\rho(x)$.

The main observation is that $\rho(x)$ has several peaks with holes around
them.  These peaks correspond with exact degeneracies of eigenvalues: they
are {\em delta} peak in the sense of distribution theory.  Unfortunately,
their heights on Fig.~\ref{fig:rho} are not representative because they
depend on the width of the histogram bin, chosen arbitrarily.  On the other
hand, they are regularly represented by the vertical steps on
Fig.~\ref{fig:cumul} but now only the main ones are visible.  Numerical
estimations of their heights are given in Sect.~\ref{sec:stat}.

The largest delta peaks are, in order of importance, at $x=0, \pm 1, \pm
\sqrt{2}$, etc. which are precisely the eigenvalues of the small trees
listed in Table~\ref{tab:spectra}: $T_1$, $T_2$, $T_3$, etc.  This has been
previously observed in several models of random graphs [Kirkpatrick and
Eggarter, 1972; Evangelou, 1983; Farkas and al., 2001].  In
Sect.~\ref{sec:existence} we show that the spectral density, in the large
$n$ limit, has an infinity of delta peaks at all eigenvalues of finite
trees and that the corresponding eigenvectors are strictly localized, in
the sense that their number of non-vanishing coordinates (i.e. the number
of nodes on which the vector does not vanish) is finite.  

Moreover we have noticed numerically that the reciprocal is true: all the
degeneracies (i.e. two or more equal values) among the complete set of
Monte-Carlo eigenvalues appear only at eigenvalues of finite trees and that
the corresponding eigenvectors are strictly localized on a small number of
nodes.  In other words, the delta peaks appear at these special values and
nowhere else.  However the heights of peaks are exponentially small with
the size of the corresponding finite subtree.  Thus on Fig.~\ref{fig:rho}
and \ref{fig:cumul}, only the largest ones are visible because the smallest
ones are drowning into Monte-Carlo fluctuations.

To summarize, our simulations indicate that the asymptotic distribution
$\rho(x)$ has two components in the large $n$ limit: a {\em discrete}
component $\rho_d(x)$ made of an infinity (but countable) of delta peaks at
all eigenvalues of finite trees associated with strictly localized
eigenvectors, and a {\em continuous} component $\rho_c(x)$ built with all
eigenvalues which are not eigenvalues of finite trees which form a
continuous support.  As explained in Sect.~\ref{sec:stat}, the total weight
of $\rho_c(x)$ is $0.80827 \pm 0.00005$.  Consequently, the total weight of
$\rho_d(x)$ is $0.19173 \pm 0.00005$.

Note that the spectral density of the trees on $n$ nodes, for any $n$, is
purely discrete because it is defined by a finite list of $n^{n-1}$
eigenvalues.  So the continuous component $\rho_c(x)$ appears only in the
large $n$ limit.  However it is possible to study $\rho_c(x)$ with
Monte-Carlo finite trees with the following procedure.  Firstly we
enumerate all ``small'' trees on $n \le 23$ nodes and their spectra are
listed.  Next, during Monte-Carlo simulations of much larger trees, we
research if the generated eigenvalues are among the spectra of small trees.
If they are, we regard them as part of delta peaks; otherwise, we regard
them as part of $\rho_c(x)$.

Of course, the threshold $n=23$ is arbitrary and dictated by the power of
our computer.  But this threshold can be reduced up to $n \approx 15$
without significant changes because the delta peaks corresponding to trees
on $n > 15$ nodes are so small that they are not visible in our
simulations.  As the heights of peaks decrease exponentially
with $n$, the effective threshold increases as the logarithm of the
computational time: then the value $n=23$ is greatly sufficient.

\begin{figure}
  \centering
  \includegraphics [width=\figwidth] {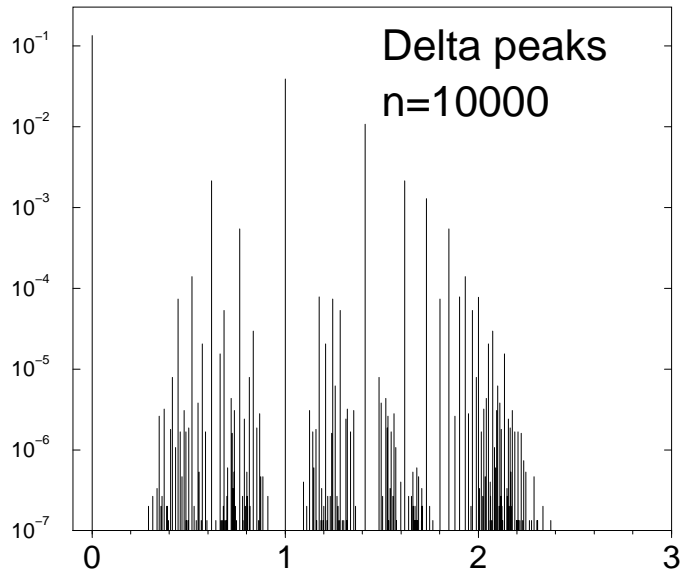}
  \caption{\em 
    Delta peaks of the spectral density: eigenvalues of Monte-Carlo trees
    on $n=10000$ nodes which are also eigenvalues of a small trees.  Because
    of its symmetry, the histogram is folded up around zero.  Note that the
  $y$-axis is logarithmic.
  } 
  \label{fig:peak}
\end{figure}

With the previous criterion, we show on Fig.~\ref{fig:peak} the discrete
spectral density $\rho_d(x)$ represented by the histogram of eigenvalues of
Monte-Carlo trees on $n=10000$ nodes which are also eigenvalues of small
trees, drawn as delta peaks with now a correct height.  As $30 \times 10^6$
eigenvalues are generated, only peaks with height bigger than $10^{-7}$ are
observed with sufficient statistics.  Because of this numerical cutoff, the
reader could believe that the support of $\rho_d(x)$ has some gaps, but it
is wrong because the eigenvalues of finite trees are a dense (but
countable) subset of real numbers.  A proof is given in
Appendix~\ref{sec:density}.

\begin{figure}
  \centering
  \includegraphics [width=\figwidth] {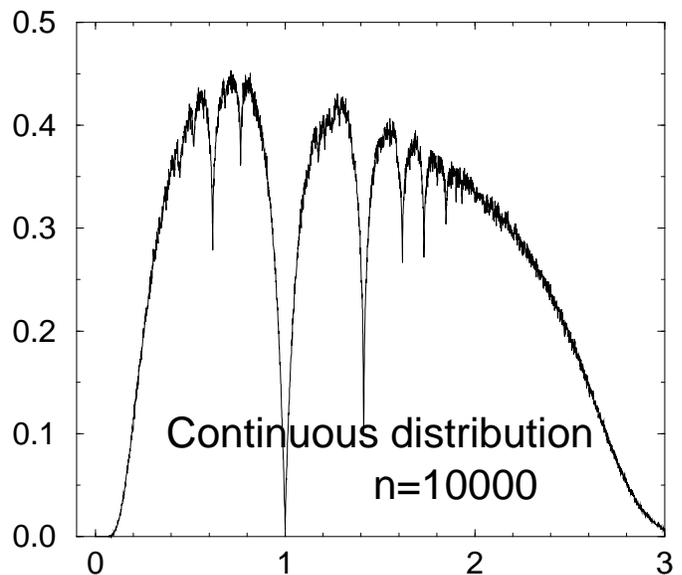}
  \caption{\em 
    Continuous component $\rho_c(x)$ of the spectral density: eigenvalues
    of Monte-Carlo trees on $n=10000$ nodes which are not eigenvalue of
    a small tree.  Because of its symmetry, the histogram is folded up
    around zero.
  } 
  \label{fig:cont}
\end{figure}

The histogram of the other eigenvalues (i.e. which are not eigenvalues of
small trees) is drawn on Fig.~\ref{fig:cont}: therefore it represents the
continuous component $\rho_c(x)$ of the spectral density.  As expected, it
looks like Fig.~\ref{fig:rho} but without peaks.  In particular,
$\rho_c(x)$ has holes at $x=0, \pm 1, \pm \sqrt{2}$, etc. which are
positions of the delta peaks.  By varying the width of the histogram bins,
$\rho_c(x)$ appears to be zero and continuous at these special values, even
if it is not really visible on Fig.~\ref{fig:cont}.  Around each special
value, the spectral density concentrates to built a delta peak with a
finite height and this concentration leaves a hole.  This effect is
proportional to the height of the peak.  Those holes remember that in
quantum mechanics, an eigenvalue with a non localized eigenvector is
repelled by the others.

By generalization, we conjecture that any delta peak makes a hole in
$\rho_c(x)$ proportional to its height.  As delta peak positions are a
dense (but countable) subset of the real numbers, $\rho_c(x)$ is a density
function, non-continuous and non-zero almost everywhere except at these
special values where it is continuous and zero.  From a mathematical point
of view, note that such ``pathological'' functions exist, for example $f(x)
= \exp(-x^2) \min_i \{ 2^i\ (x-\lambda_i)^2 \}$ where the
$(\lambda_i)_{i\ge 1}$ are the special points.  Remark that there is no
contradiction with the word {\em continuous}: in the sense of the measure
theory, $\rho_c(x)$ is a continuous distribution because its support is
continuous or equivalently its integral $\int_{-\infty}^x \rho_c(y)\ dy$ is
a continuous function.  On the contrary, the discrete distribution $\rho_d$
has a discrete support and its integral has discontinuities (or jumps) for
all delta peaks.

A sceptical reader could object that $\rho_c(x)$ could have a {\em
singular} continuous component, i.e. a measure everywhere zero excepts at a
non-countable set with Lebesgue measure zero.  A famous example on
$[0,1]$ is the Cantor measure on the Cantor set.  In this measure, the sum
of the width of the gaps is 1 and all the measure is concentrated on the
Cantor set.  As this set has Lebesgue measure zero, the density is infinite
on its points.  But it is not a discrete distribution (with delta peaks)
because this set is non-countable then the measure of any point remains
always strictly zero.  Equivalently, the integral is a continuous function.

Our numerical analysis can not determine if $\rho_c(x)$ contains or not a
singular part because Monte-Carlo fluctuations prevent from distinguishing
between a singular measure and a measure with a pathological density
function.  But we think that $\rho_c(x)$ vanishes only on the delta peaks
positions, which are a set with Lebesgue measure zero, unlike the Cantor
measure.  Then the density remains finite everywhere, even if it is a
pathological function.  As we do not succeed to imagine a mechanism which
concentrates the spectral density on a non-countable set with Lebesgue
measure zero, we believe that $\rho_c(x)$ is not singular.

To summarize this section, the spectral density of large random trees have
an infinity of delta peaks at all eigenvalues of finite trees, and a
pathological density function vanishing at all peak positions.

\section{Delta peaks in the spectral density}

\subsection{Existence of delta peaks}
\label{sec:existence}

To show that a delta peak appears in the spectral density of large random
trees at any eigenvalue of finite tree, we will adapt arguments of [Bauer
and Golinelli, 2001a, p 322] used for the same property in random graphs.
First, we show that in a large tree on $n$ nodes, the number of branches
isomorphic to any finite tree $T$ is proportional to $n$.  Consequently
delta peaks appear in the spectral density because the multiplicity of
eigenvalues of $T$ with eigenvectors strictly localized on branches is also
${\cal O}(n)$ in large trees.

We now introduce two definitions.  A {\em rooted tree} $(T,r)$ is a tree
$T$ where the node $r$ is marked (the {\em root}).  A {\em branch} $(B,r)$
at a node $r$ in a tree $T$ is a rooted subtree of $T$ with no edges
between nodes of $B \setminus \{r\}$ and nodes of $T \setminus B$. Remark
that it is not the usual definition.

Let us consider any rooted tree $(T,r)$ on $m$ nodes.  The number of
branches isomorphic to $(T,r)$ among all labeled trees on $n$ nodes is
\begin{equation}
  B(n) = \frac{n!}{S\ (n-m)!}\ (n-m+1)^{(n-m-1)},
  \label{eq:bn}
\end{equation}
because the first factor counts the number of ways of labeling the branch
($S$ is the symmetry factor of $(T,r)$), and the second factor counts the
choices of the rest of the tree, i.e. $(n-m)$ nodes, plus $r$.
Consequently, a tree on $n$ nodes has in average $b(n) = B(n)/n^{n-2}$
branches isomorphic to $(T,r)$; in the large $n$ limit,
\begin{equation}
  \label{eq:btn}
  b(n) \sim n \ \frac{ e^{-(m-1)}}{S}.
\end{equation}
For example, $m=2$ gives $n/e$ as the number of leaves in average in a
large tree.  Moreover, by considering more complex patterns with two
branches, it can be proved that the variance of this number of branches per
tree is ${\cal O}(n)$ because contributions in ${\cal O}(n^2)$ are
canceled.  Then this number is ``self-averaging'': in the large $n$ limit,
for almost all trees, the number of branches isomorphic to $(T,r)$ is $n
e^{-(m-1)}/S $ plus ${\cal O}(\sqrt{n})$ fluctuations.  The amplitude of
theses fluctuations has no simple formula because it depends not only on
the size $m$ but also on the shape of $T$.

Let $U$ be a tree with a branch $(T,r)$ at a node $r$; let $(\lambda, V)$
be an eigenmode of $T$: if $V_r=0$, then $\lambda$ is eigenvalue of $U$.
Indeed the vector $W$ on $U$ which extends $V$ with coordinates zero
outside $T$ is also eigenvector with same eigenvalue $\lambda$,
independently of the size and the shape of the rest of $U$.  This shows
that $\lambda$ can be found among spectra of large trees, with an
eigenvector strictly localized on any branch (i.e. strictly zero outside)
isomorphic to $(T,r)$.  As the number of such branches in a large tree is
${\cal O}(n)$, the multiplicity of $\lambda$ is also ${\cal O}(n)$: this
gives a delta peak at $\lambda$ in the spectral density with strictly
localized eigenvectors.

\begin{figure}
  \centering
  \includegraphics{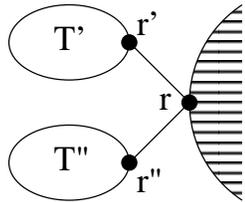}
  \caption{\em 
    Construction of the tree $D$ by duplication of the initial tree $T$
    (see text). As the eigenvector vanishes on $r$, $D$ can be ``grafted''
    on a larger tree (shaded area) for which the eigenvector is extended with
    coordinates zero.
  }
  \label{fig:double}
\end{figure}

Of course, this is true for any eigenvalue of $T$ if the corresponding
eigenvector $V$ vanishes on (at least) one node, by choosing this node as
branching node $r$.  In case all the coordinates $V_i \ne 0$, we transform
$T$ (with $m$ nodes) into a larger tree $D$ on $2m+1$ nodes, with the same
eigenvalue $\lambda$ and now an eigenvector $W$ with a coordinate $W_r=0$.
As shown on Fig.~\ref{fig:double}, $D$ is made of two copies $T'$ and $T''$
of $T$, plus one node $r$, plus two edges $(r,r')$ and $(r,r'')$ where $r'$
and $r''$ are chosen among the nodes of $T'$ and $T''$ respectively (and
not necessarily symmetric).  We note $a' = V_{r'}$ and $a'' = V_{r''}$
(following the hypothesis, $a'$ and $a''\ne 0$).  Let $W$ be the vector on
the nodes of $D$ where the restriction on $T'$ is $a''V$, the restriction
on $T''$ is $-a'V$ and $W_r=0$.  As $W_{r'} + W_{r''} = 0$, the eigenvalue
Eq.~(\ref{eq:lambda}) is satisfied for $r$, then $(\lambda, W)$ is
effectively an eigenmode of $D$, but with $W_r = 0$.  Now, we use the
previous arguments with $D$ in place of $T$ to show that a delta peak
appears at $\lambda$ in spectra of large trees.  Of course this
construction is possible but useless if the initial vector has a coordinate
zero.

\label{sec:generD}
Remark that this construction can be generalized by many ways.  The trees
$T'$ and $T''$ are not necessarily isomorphic: the only constraint is that
they must have $\lambda$ in common in their spectra.  In particular, this
construction can be iterated, by using $D$ or $U$ in place of $T'$ or
$T''$.  Moreover, the transformation can involve $k \ge 2$ trees and an
additional node $r$ connected to $k$ nodes $(r_i)_{i=1,k}$ (one per tree)
with coordinates non-zero.  The multiplicity of $\lambda$ is $k-1$ because
the coefficients $a_i$ must satisfy the eigenvalue condition on $r$:
$\sum_i a_i V_{r_i} = 0$.

The previous argument shows that the delta peaks do not vanish in the large
$n$-limit but Eq.~(\ref{eq:btn}) gives only a coarse approximation of their
height for many reasons.  First, several finite trees with the same $\lambda$
contribute to the same peak (see Appendix~\ref{sec:pure}) and
Eq.~(\ref{eq:btn}) shows that the main contributions come from trees with
shape like $D$ (see Fig.~\ref{fig:double}) and two components $T'$ and
$T''$ with minimal size $m$.  This contributions is $s\: e^{-2m}$ where
$s$ counts choices of $T'$ and $T''$ and choices of roots, without
forgetting symmetry factors.  Secondly, in the case where the construction
of the above paragraphs is done with $k \ge 3$ trees, the multiplicity of
$\lambda$ is $k-1$ but the tree $D$ is counted $k(k-1)/2$ times in
Eq.~(\ref{eq:bn}).  So we must use inclusion-exclusion formula: this 
gives a ${\cal O}(e^{-3m})$ correction.

Because of these remarks, we do not know analytical formula for the heights
of the delta peaks except at $\lambda=0$ [Bauer and Golinelli, 2000] which
corresponds to the tree on $m=1$ node.  However we keep the main result: in
the large random tree limit, the multiplicity of any eigenvalue of finite
tree is ${\cal O}(n)$ and a delta peaks appears with a self-averaging
height.

\subsection{Statistics of delta peaks}
\label{sec:stat}

In this section, we give numerical estimations for the heights of delta
peak in the spectral density of large random trees.  We have explained in
previous sections that these peaks correspond to eigenvalues of finite
trees.  As their number increases exponentially with their size, it would
be cumbersome to give a list of heights for each eigenvalue.  So we prefer
to group eigenvalues by {\em order}.

Let $\lambda$ be an eigenvalue of a tree on $n$ nodes.  We define the {\em
order} of $\lambda$ as the size $d$ of the smallest tree(s) with eigenvalue
$\lambda$.  Of course, $d \le n$.  We define ${\cal L}_d$ as the set of
eigenvalues of order $d$: it is the finite set of eigenvalues of trees on
$d$ nodes but not in ${\cal L}_{d'}$ with $d'<d$.

In principle, ${\cal L}_d$ can be explicitly written by enumerating all
spectra of trees on $d' \le d$ nodes, as done on Table~\ref{tab:spectra}.
As the spectrum of the trivial tree $T_1$ on one node is $\{0\}$, then
${\cal L}_1 = \{0\}$.  For $d=2$, the tree $T_2$ gives ${\cal L}_2 = \{-1,
1\}$.  For $d=3$, the spectrum of $T_3$ is $\{-\sqrt{2}, 0, \sqrt{2}\}$.
As 0 is already of order 1, then ${\cal L}_3 = \{-\sqrt{2}, \sqrt{2}\}$.
The following sets are ${\cal L}_4 = \pm \{(\sqrt{5} \pm 1)/2, \sqrt{3} \}$,
${\cal L}_5 = \pm \{ \sqrt{2 \pm \sqrt{2}}, 2 \}$, etc.  Generally, as all
spectra are symmetric with respect to zero, the ${\cal L}_d$ are also
symmetric.

An eigenvalue of order $d$ is by definition a root of a characteristic
polynomial, which is an integer polynomial of degree $d$.  But this
polynomial can often be factorized, for example if $d$ is odd (because 0 is
eigenvalue by symmetry) or if the tree has some symmetries [Cvetkovic et
al., 1995, ch 4].  So an eigenvalue of order $d$ is an algebraic number of
degree $d'$ with $d'\le d$.  It would be interesting to be able to
determine the order of a given eigenvalue of a tree $T$ without have to
enumerate all the trees smaller than $T$.  However, we have only a partial
answer expounded in Appendix~\ref{sec:pure}.

We define ${\cal F}_d$ as the sum of the heights of delta peaks
corresponding to eigenvalues of order $d$, in the large random tree limit.
In other words, ${\cal F}_d$ is the fraction of the spectrum of a large
random tree occupied by eigenvalues of order $d$.

\begin{table*}  \centering
  \begin{tabular}{r|l|lll|l}
    $d$ & $n = 23$ & $n = 100$ & 1000 & 10000 & $n = \infty$ \\
  \hline
 1 & 0.14371 &  0.13643  &  0.13456  &  0.13438(6)  & 0.13433(3) \\
 2 & 0.05187 &  0.04148  &  0.03906  &  0.03883(5)  & 0.03879(3) \\
 3 & 0.01937 &  0.01228  &  0.01086  &  0.01076(3)  & 0.01074(2) \\
 4 & 0.01486 &  0.00660  &  0.00564  &  0.00554(3)  & 0.00555(2) \\
 5 & 0.00486 &  0.00147  &  0.00120  &  0.00117(1)  & 0.00116(1) \\
 6 & 0.00456 &  0.00092  &  0.00072  &  0.00072(1)  & 0.00072(1) \\
 7 & 0.00251 &  0.00033  &  0.00026  &  0.00027(1)  & 0.00026(1) \\
 8 & 0.00210 &  0.000148 &  0.000115 &  0.000116(5) & 0.000115(3) \\
 9 & 0.00125 &  0.000043 &  0.000035 &  0.000032(3) & 0.000031(2) \\
10 & 0.00131 &  0.000024 &  0.000012 &  0.000015(2) & 0.000015(1) \\
11 & 0.00130 &  0.000010 &  0.000006 &  0.000005(1) & 0.000005(1) \\
12 & 0.00121 &  0.000003 &  0.000002 &  0.000004(1) & 0.0000025(5) \\
  \hline
cc &         &  0.80024 &  0.80752 &  0.80816(8) & 0.80827(5) \\
  \end{tabular}
  \caption{\em
    The fraction ${\cal F}_d$ of the spectrum of a random tree on $n$ nodes
    occupied by eigenvalues of order $d$.  The column $n=23$ gives exact
    results.  The columns $n=$ 100, 1000 and 10000 give Monte-Carlo
    results; the one-sigma standard error (number inside parentheses) is
    given only for $n=$ 10000 because it is on the same amplitude for the
    other $n$.  The column $n=\infty$ gives extrapolations (see text).  The
    row labeled ``cc'' is the complement, i.e. the weight of the continuous
    component in the spectral density.
  }
  \label{tab:resu}
\end{table*}

For $d=1$ (i.e. $\lambda=0$) we know [Bauer and Golinelli, 2000] that
${\cal F}_1 = 2W-1 = 0.134286\dots$, where $W = 0.567143\dots$ is the root
of $e^{-x}=x$.  However for $d \ge 2$, we have no analytic formula and
numerical results are given in Table~\ref{tab:resu}.  Results for $n=23$
are obtained by exact enumeration, described in Sect.~\ref{sec:enu}.  For
larger $n$ we use the Pr\"ufer Monte-Carlo method described in
Sect.~\ref{sec:pmc} by averaging over $m = 30 \times 10^6$ eigenvalues for
each size $n$.

Moreover we know [Bauer and Golinelli, 2000] that ${\cal F}_1$ for a finite
size $n$ has an asymptotic expansion in powers of $1/n$.  We observe
numerically the same behavior for $d \ge 2$.  So we extrapolate for $n =
\infty$ with least-square fits with linear or quadratic function of $1/n$.
For these fits, we include Monte-Carlo results for $n = 30, 50, 100, 200,
500, 1000, 2000, 5000$ and $10000$ and the standard errors for the
extrapolations are smaller than the ones of the measures, as shown on
Table~\ref{tab:resu}.  By simulating trees on $n=10001$ nodes, we have also
verified that the results do not show perceptible parity effects.

We also observe numerically that, for a given $d$, the variance of ${\cal
F}_d$ for random trees on $n$ nodes is proportional to $1/n$: in other
words, the ${\cal F}_d$ are self-averaging for large trees.  The opposite
would be very surprising because the eigenvectors corresponding to
eigenvalue of finite order are strictly localized on finite patterns: the
appearances of these patterns look like independent events on a large
random tree.  A consequence is that the standard error of ${\cal F}_d$ is
$O(1/\sqrt{m})$ (for a given $d$) where $m$ is the total number of
generated eigenvalues, independently of the size $n$.

We are also interested by the localization of eigenvectors.  For an
eigenvector $V$ corresponding to an eigenvalue of order $d$, let $l$ be
the number of coordinates (or nodes) where $V_i \ne 0$.  We observed that
most of them are localized on $l = 2d$ nodes: they correspond to the
construction shown on Fig.~\ref{fig:double} with $T'$ and $T''$ as
smallest as possible.

As explained in Sect.~\ref{sec:existence} and Appendix~\ref{sec:pure},
eigenvectors with arbitrary large $l$ can be built.  But the probability of
appearance of large patterns decreases exponentially with their size.  So
in our Monte-Carlo simulations, a few of eigenvectors have an extension $l
> 2d$ without ever exceeding $4d$.  In principle, with more important
simulations, it would be possible to observe larger (and rarer) localized
eigenvectors.

In particular we do not remark a delocalization phenomenon, as observed
[Bauer and Golinelli 2001] in the Erd\"os-R\'enyi random graph model for
eigenvectors with $\lambda=0$, when the average {\em effective}
connectivity is between $2.093\dots$ and 3.312\dots  Note that in the
random tree model, the average connectivity is by definition fixed to 2.

\section{Conclusion}

In this work we have studied the spectral density of the adjacency
matrix of random labeled trees, as a model of a hopping particle on a
connected graph.  In the limit where the number $n$ of nodes is large, our
main results are:
\begin{itemize}
\vspace{-1ex} \item 
The averaged spectral density converges to an
asymptotic distribution $\rho(x)$ with a $O(1/n)$ behavior.  Moreover the
spectral density is self-averaging, i.e. the spectral density of a given
tree is $\rho(x)$ almost surely.

\vspace{-1ex} \item 
At all eigenvalues of finite trees, $\rho(x)$ has a delta peak.  Its height
decreases exponentially with the size of the corresponding finite tree.
Except for the $\lambda=0$ peak, we do not know how to calculate
analytically these heights, but numerical estimations are given.  The total
weight of these peaks is $0.19173 \pm 0.00005$.

\vspace{-1ex} \item 
The rest of $\rho(x)$ is a distribution given by a density function which
vanishes at each position of delta peak.  As these positions form a dense
set among real numbers, this function is almost everywhere discontinuous.

\vspace{-1ex} \item 
The eigenvectors corresponding to eigenvalues of delta peaks are
strictly localized, i.e. they vanish everywhere except on a finite number
of nodes.  On the other hand, the other eigenvectors are not strictly
localized.
\end{itemize}

\vspace{-1ex} 

It seems difficult to extend the analytical calculation [Bauer and
Golinelli, 2000] of the height of the $\lambda=0$ peak to all peaks because
it is specific to this particular eigenvalue: there is only one ``pure''
tree for $\lambda=0$ and moreover this tree is the trivial tree on one
node.  Then the enumeration of composite trees with $\lambda=0$ is
possible.  In contrast, each other peak corresponds to an infinite family
of pure trees, which are not trivial.

We think that these results are shared by many models of
graphs.  For any eigenvalue of finite graph $g$ with eigenvector $V$, a
delta peak appears in the spectral density of a large graph $G$ on $n$
nodes if the number of induced subgraphs of $G$ isomorphic to $g$ (and
connected to the rest of $G$ only by nodes on which $V$ vanishes) is
$O(n)$.  Of course the height of the delta peak depends on the details of
the model.  Generally, the highest ones correspond to the smallest $g$
and their contribution gives an important fraction of the total weight of
delta peaks.

Furthermore the existence of a pathological density function for the
continuous part of the spectral density is not specific to the random
trees.  As it corresponds to eigenvectors which are not strictly localized
and orthogonal to eigenvectors of the delta peaks, it is expected that the
density function has a depression around each delta peak.  Moreover we
think that these conclusions are also valid for other kind of Hamiltonian,
for example the Laplacian, or graphs with weighted links.  

\appendix 

\section{Symmetry of the spectrum}

\label{sec:symspec}

In this appendix, we show that the bipartition of any tree induces that its
spectrum is symmetric with respect to zero.  Moreover it allows to reduce
the computation time of the spectrum by a factor 8.

Let $T$ be a tree on $n$ nodes.  The set of nodes can be partitioned into
two subsets, $P$ and $Q$, of sizes $p$ and $q$ (with $p+q=n$), so that all
edges link a $P$ node with a $Q$ node.  We consider the spectrum
$(\lambda_i)_{i=1,n}$ of the symmetric adjacency matrix $A$ of $T$.  In
this notation, degenerate eigenvalues correspond to several indices $i$.
Note that all eigenvalues are real.  First, without changing the spectrum,
we permute the labels of the nodes of $T$ so that $P$ nodes are labeled by
$(1,2, \dots, p)$ and $Q$ nodes are labeled by $(p+1, p+2, \dots, n)$.  In
this basis, the rows and columns of $A$ are permuted and $A$ has now a
$(p+q) \times (p+q)$ block shape
\begin{equation}
  A =  \left( \begin{array}{cc} 0 & R^T \\ R & 0 \end{array} \right)
\end{equation}
where $R$ is a rectangular $q\times p$ block and $R^T$ is the transpose of
$R$.  Then, 
\begin{equation}
  A^2 =  \left( \begin{array}{cc}  R^T R & 0 \\ 
                                   0     & R R^T \end{array} \right)
\end{equation}
is block diagonal: its spectrum $(\lambda_i^2)_{i=1,n}$ is the union of the
spectra of the square $p \times p$ block $R^T R$ and of the square $q
\times q$ block $R R^T$.  We will show that it is sufficient to diagonalize
only one block, $R^T R$ for example.

Let $V$ be an eigenvector of $R^T R$ ($V$ is $p$-dimensional) with
eigenvalue $\lambda^2$, then $R^T R V= \lambda^2 V$.  We assume that
$\lambda \ne 0$.  Then $RV$ is a ($q$-dimensional) eigenvector of the other
block $RR^T$ with the same eigenvalue $\lambda^2$, because $RR^TRV =
R(\lambda^2 V) = \lambda^2 RV$.  The two $n$-dimensional vectors
\begin{equation}
  U_{\pm} = \left( \begin{array}{c} \pm \lambda V \\ 
                                         RV          \end{array} \right)
\end{equation}
are eigenvectors of $A$ with eigenvalues $\pm\lambda$ because 
\begin{equation}
  A U_{\pm} =
  \left(\begin{array}{c} R^TRV \\ \pm \lambda RV \end{array} \right) =
  \left(\begin{array}{c} \lambda^2 V \\ \pm \lambda RV \end{array} \right)
  = \pm \lambda U_{\pm}
\end{equation}
If $\lambda^2$ is a degenerate eigenvalue of $R^TR$, this work can be done
with any corresponding eigenvector.

To resume, any eigenvalue $\lambda^2 > 0$ of one block of $A^2$ is
associated to an eigenvalue $\lambda^2$ of the other block and a couple
$\pm \lambda$ of $A$.  The rest is made of eigenvalues 0, with $m(A) =
m(R^TR) + m(RR^T)= 2m(R^TR) + q - p$ by noting $m(\cdot)$ its multiplicity.
Then the spectrum of $A$ is symmetric with respect to zero.  Moreover
eigenvalues (and eigenvectors) of $A$ can be deduced from the
diagonalization of only one block.

From a numerical point of view, the complete diagonalization of a $n \times
n$ full matrix needs a time ${\cal O}(n^3)$.  As $\min(p,q) \le n/2$, it is
at least 8 times faster to diagonalize the smallest block, $RR^T$ or
$R^TR$, instead of $A$.  In the large $n$ limit, $p \sim q \sim n/2$ for
almost all trees, so the asymptotic factor is 8.

Remark that the characteristic polynomial of $A$ can by computed in ${\cal
O}(n^2)$ by using a recursion procedure with the characteristic polynomials
of the subtrees obtained by deleting one node.  However the computation of
eigenvalues from the knowledge of the characteristic polynomial is
numerically very instable, so we prefer diagonalize $A$ with Lapack
routines for symmetric full matrix.

\section{Density of the eigenvalues of finite trees}
\label{sec:density}

In this appendix, we show that the set ${\cal L}$ of eigenvalues of finite
trees is a dense subset of real numbers.  Remark that by definition ${\cal
L}$ is countable because finite trees are countable.  We proceed in two
steps: firstly the eigenvalues of linear trees are a dense subset of
$[-2,2]$.  Secondly the eigenvalues of linear trees decorated with bunches
of $k$ leaves are a dense subset of $\pm[\sqrt{k+1}-1,\sqrt{k+1}+1]$.  As
the union of all these intervals covers the real numbers, ${\cal L}$ is
everywhere dense.

\begin{figure}
  \centering
  \includegraphics{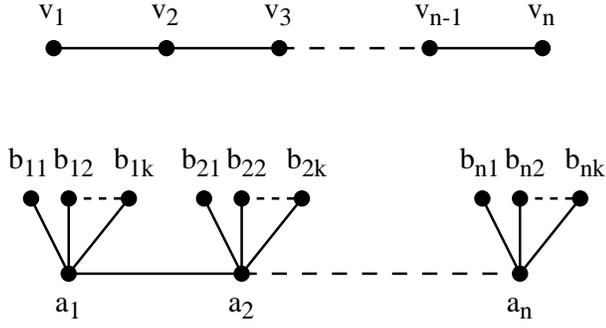}
  \caption{\em 
    Top: a linear tree of length $n$.  Bottom: a linear tree of length $n$
    decorated with bunches of $k$ leaves.  
  }
  \label{fig:linear}
\end{figure}

For a linear tree of length $n$, pictured on Fig.~\ref{fig:linear},
$\lambda$ is eigenvalue with eigenvector $V = \{v_i\}_{i=1,n}$ if
\begin{equation}
  \lambda\ v_i = v_{i-1} + v_{i+1} \ \ \ \ 
  (i = 1, 2, \dots,n),
  \label{eq:vvv}
\end{equation}
with the conventions $v_0 = v_{n+1} = 0$.
The $n$ solutions are
\begin{equation}
  \lambda_p = 2 \cos \left( \frac{p \pi}{n+1} \right), \ \
  v_{i,p} =   \sin \left( \frac{i p \pi}{n+1} \right) 
\end{equation}
for $p = 1, 2, \dots, n$.
Then the union of spectra of finite linear trees is dense in [-2,2].

Now we consider a linear tree of length $n$ decorated with bunches of $k
\ge 1$ leaves: it is made of a linear backbone $\{a_i\}_{i=1,n}$, and for
each $i$, $k$ nodes $\{b_{i,j}\}_{j=1,k}$ are connected to $a_i$.  See
Fig.~\ref{fig:linear}.  Its size is $(k+1)n$. 

For an eigenvalue $\lambda$, we note $a_i$ and $b_{i,j}$ the
coordinates of the eigenvector on the respective nodes.
The eigenvalue equations are
\begin{eqnarray}
  \lambda\ b_{i,j} &=& a_i\ \ \ (i = 1, 2, \dots,n; \ \ j = 1, 2, \dots, k)\\
  \lambda\ a_i     &=& a_{i-1} + a_{i+1} + \sum_{j=1}^k b_{i,j}
  \ \ \ (i = 1, 2, \dots,n)\ \ \ \ 
\end{eqnarray}
with the conventions $a_0 = a_{n+1} = 0$.  
If $\lambda = 0$, then
\begin{equation}
  \sum_{j=1}^k b_{i,j} = a_i = 0 \ \ \ (i = 1, 2, \dots,n),
\end{equation}
and the multiplicity of the eigenvalue 0 is $(k-1)n$.
For $\lambda \ne 0$, 
\begin{equation}
  b_{i,j} = a_i / \lambda,
  \ \ \ (\lambda - k/\lambda)\ a_i = a_{i-1} + a_{i+1}.
  \label{eq:lkl}
\end{equation}
Then the equation for $\mu = \lambda - k/\lambda$ is similar to
Eq.~(\ref{eq:vvv}).  There are $2n$ non-zero eigenvalues given by
\begin{eqnarray}
  \lambda_p^{(\pm)} &=&
                  \frac{1}{2} \left(\mu_p \pm \sqrt{\mu_p^2 + 4k}\,\right), \\
  \mu_p &=& 2 \cos \left( \frac{p \pi}{n+1} \right) \ \ \ (p = 1, 2, \dots, n).
  \label{eq:mu}
\end{eqnarray}
As the set of $\mu_p$ for $n\ge 1$ is dense in $[-2,2]$, then the set of
$\lambda_p^{(+)}$ is dense in $[\sqrt{k+1}-1, \sqrt{k+1}+1]$ and the set of
$\lambda_p^{(-)}$ is dense in $[-\sqrt{k+1}-1, -\sqrt{k+1}+1]$.

The union of these intervals for $k \ge 1$, plus the interval $[-2,2]$
given by linear trees, covers all the real numbers.  So the trees described
above are sufficient to prove that eigenvalues of finite trees are a dense
subset of real numbers.

\begin{figure}
  \centering
  \includegraphics{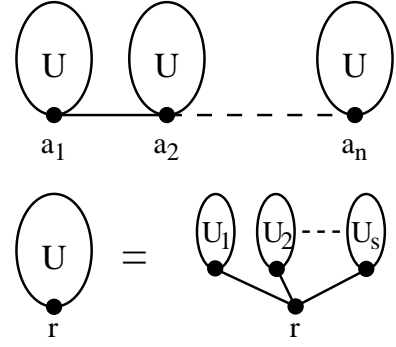}
  \caption{\em 
    Top: a linear tree of length $n$ decorated with isomorphic trees $U$.
    Bottom: decomposition of $U$.
  }
  \label{fig:lingen}
\end{figure}

We remark that the same calculation can be done with other decoration
patterns.  Let us consider now a tree with a linear backbone
$\{a_i\}_{i=1,n}$, and for each $i$, a rooted tree $(U,r)$ is grafted by
merging $r$ and $a_i$, as pictured on Fig~\ref{fig:lingen}.  Then,
Eq.~(\ref{eq:lkl}) is generalized by
\begin{equation}
  f(\lambda) \ a_i = a_{i-1} + a_{i+1},
  \label{eq:fl}
\end{equation}
where the function $f(x)$ is defined by
\begin{equation}
  f(x) = \frac{\chi_U(x)}{\prod_{j=1}^s \chi_{U_j}(x)}
\end{equation}
by noting $\chi_T(x)$ the characteristic polynomial of the adjacency matrix
of a tree $T$, and $U_1$, $U_2$, \dots $U_s$ the subtrees of $U$ obtained by
removing its root $r$ (see Fig~\ref{fig:lingen}).  The eigenvalues are the set of $f^{-1}(\mu_p)$,
solutions of $f(x)=\mu_p$ and they are dense in $f^{-1}([-2,2])$.  

That remembers the band theory in solid state physics.  We look at $U$ as
an atom: an eigenvalue $f^{-1}(0)$, solution of $\chi_U(x) = 0$, is an
energy level.  The linear tree decorated with $U$ is like a crystal, and
the terms $a_{i\pm1}$ in Eq.~(\ref{eq:fl}) mean that electrons can hope
from an atom to its neighbors.  Then each atomic level $f^{-1}(0)$ is
stretched and forms a band $f^{-1}(2\cos(k))$ where $k$ is the wave vector.

\section{Large pure trees}
\label{sec:pure}

For any finite tree $T$ with eigenvalue $\lambda$, we have seen in
Sect.~\ref{sec:generD} that it is possible to build arbitrarily large trees
with same eigenvalue $\lambda$ by connecting many copies of $T$ with
additional nodes with coordinates zero.  In this Appendix, we will show
that arbitrarily large trees with eigenvectors without coordinates zero can
be obtained as well.

Let $T$ be any tree with eigenvalue $\lambda$ and eigenvector $V$.  If all
coordinates $V_i \ne 0$ , we say that $T$ is $\lambda$-{\em pure}.  If $T$
is $\lambda$-pure, it can be proved that $\lambda$ is not degenerate in the
spectrum of $T$.  Then the $\lambda$-purity is really a property of the
tree, and it does not depend on a particular choice of eigenvector.  By
noting $\lambda_m$ the largest eigenvalue of $T$, remark that $T$ is
$\lambda_m$-pure because the Perron-Frobenius theorem assures that all
coordinates of the corresponding eigenvector are all positive.

If $T$ is not $\lambda$-pure, it is composite.  By deleting any node $i$
with $V_i = 0$, $T$ is split into a forest of one or more subtree(s); on
each subtree, the restriction of $V$ is again an eigenvector with
eigenvalue $\lambda$.  Then by deleting all nodes with coordinates 0, the
rest is a forest of $\lambda$-pure subtrees.  A consequence is that the
order of $\lambda$ is less or equal to the size of the smallest pure
subtree.

Naturally we wish enumerate $\lambda$-pure trees.  The case $\lambda=0$ can
be easily solve: for any leaf $i$, its neighbor $j$ has $V_j = \lambda V_i
= 0$.  So there is only one 0-pure tree: the trivial tree on one node.  On
the contrary, if $\lambda \ne 0$, we will show that there is an infinity of
$\lambda$-pure trees.

Many processes can built an arbitrary large $\lambda$-pure tree $U$ by
using several $\lambda$-pure trees $T_i$ (not necessarily isomorphic) and
additional nodes.  Some processes are general: they apply to all trees
$T_i$.  But other processes are specialized to trees with special patterns
or special $\lambda$ (for example integer $\lambda$).  There is an infinity
of kind of processes and we have not found a unified way to present them.
So we content ourself to give some examples.

As it is cumbersome to explain the details with words, we show graphical
representations of the resulting tree $U$ with its eigenvector $W$ by using
the following rules.  A triangle represents a $\lambda$-pure tree $T_i$ and
the restriction of $W$ on $T_i$ is the eigenvector of $T_i$ with eigenvalue
$\lambda$.  When a node $r$ is labeled by $x$, it means that $W_r = x$.  If
$r$ is a vertex of a triangle $T_i$, the eigenvector of $T_i$ is scaled
in order that $W_r = x$.  Remark that it is always possible because $T_i$
is pure, i.e. without coordinate zero.  When a node $r$ is a common vertex
of triangles $T_i$ and $T_j$, it means that a node $r_i$ of $T_i$ is merged
with a node $r_j$ of $T_j$.  The neighbors of $r$ are the ones of $r_i$,
plus the ones of $r_j$.  Of course, all these processes are done in order
that Eq.~(\ref{eq:lambda}) is satisfied for $W$ on every nodes.

\begin{figure}
  \centering
  \includegraphics{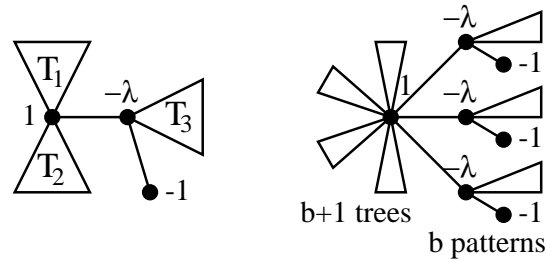}
  \caption{\em  
    Two processes to built a large ``pure'' tree with any eigenvalue of
    finite tree and an eigenvector without coordinate zero.  Left: the
    ``2+1'' process.  Right: the ``b+1,b'' process.
  }
  \label{fig:p21}
\end{figure}

The simplest general process, called ``2+1'', is shown on
Fig.~\ref{fig:p21}.  It uses three $\lambda$-pure trees $T_{\{1,2,3\}}$
with one merge and one additional node.  So $\mbox{Size}(U) = \sum_i
\mbox{Size}(T_i)$.  Remark that $U$ is pure because $\lambda \ne 0$.  The
reader can verify that Eq.~(\ref{eq:lambda}) is satisfied on every nodes of
$U$.  For example, let us consider the common node $r$ between $T_1$ and
$T_2$ with $W_r=1$: the neighbors $j_1$ of $r$ inside the tree $T_1$ give
$\sum_{j_1} W_{j_1} = \lambda W_r$, because the restriction of $W$ on $T_1$
satisfies Eq.~(\ref{eq:lambda}).  And the same for the neighbors $j_2$
inside the tree $T_2$. As $r$ is also linked with a node of $T_3$ with
coordinate $-\lambda$, then all neighbors $j$ of $r$ gives $\sum_j W_j =
\lambda W_r$.  This process can be iterated by using $U$ in place of $T_i$
to obtain arbitrary large $\lambda$-pure trees.

A generalization is the ``b+1,b'' process described on Fig.~\ref{fig:p21},
with $b+1$ pure trees with a common node linked to $b$ patterns made of one
pure tree and an additional node.  Here again $\mbox{Size}(U) = \sum_i
\mbox{Size}(T_i)$.

\begin{figure}
  \centering
  \includegraphics{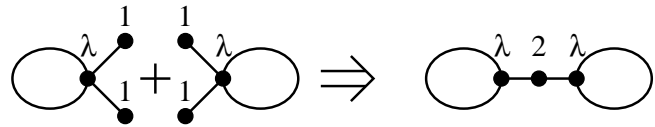}
  \caption{\em 
    A process to built a larger ``pure'' tree with two pure trees by
    merging two couples of leaves.
  }
  \label{fig:p2n3}
\end{figure}

It is easy to imagine other processes of this kind with more complex
``bridges'' between pure trees.  But we prefer to describe on
Fig.~\ref{fig:p2n3} a process which exploits a specific pattern.  It uses
two pure trees with a couple of leaves linked to the same node: $U$ is
built by merging these four leaves into a single node with a doubled
coordinate.  Then $\mbox{Size}(U) = \mbox{Size}(T_1) + \mbox{Size}(T_2) -
3$.  This process can be generalized with $k$ trees with bunches of $k$
leaves, by merging the $k^2$ leaves into a single node with a coordinate
multiplied by $k$.  Once again, more complex processes can be imagined.

To calculate analytically the height of delta peaks in the spectral
density, a method would be to enumerate the trees with eigenvalue $\lambda$
by first enumerating pure trees, and then by counting all combinations of
pure trees into composite trees.  Unfortunately, these few examples above
show that there is a great variety of pure trees, and we have no method to
do a full listing.  The only case [Bauer and Golinelli, 2000] for which
analytical results are know is for $\lambda = 0$ thanks to two facts: this
is only one 0-pure tree which has one node, and then the count of
combinations is done by a method specific to this trivial tree.

\section*{Acknowledgments}

We thank M. Bauer, J. Houdayer and K. Mallick for stimulating discussions
and a careful reading of the manuscript.

\section*{References}

\begin{itemize}

\item M. Bauer, 2000, internal report.

\item M. Bauer and O. Golinelli, 2000
   {\em On the kernel of tree incidence matrices},
   J. Integer Sequences  {\bf 3}, Article 00.1.4,
   arXiv:cond-mat/0003049. 

\item M. Bauer and O. Golinelli, 2001,
  {\em Exactly solvable model with two conductor-insulator transitions
       driven by impurities}, 
  Phys. Rev. Lett. {\bf 86}, 2621-2624,
  arXiv:cond-mat/0006472.

\item M. Bauer and O. Golinelli, 2001a,
  {\em Random incidence matrices: moments of the spectral density},
  J. Stat. Phys. {\bf 103}, 301-337,
  arXiv:cond-mat/0007127.

\item G. Biroli and R. Monasson, 1999,
   {\em A single defect approximation for localized states on random lattices},
   J. Phys. A: Math. Gen. {\bf 32}, L255-L261.

\item D.M. Cvetkovic, M. Doob, and H. Sachs, 1995,
   {\em Spectra of graphs}, 3rd ed., Barth, Heidelberg, Leipzig. 

\item P. Erd\"os and A. R\'enyi, 1960,
   {\em On the evolution of random graphs}, 
   Publ. Math. Inst. Hungar. Acad. Sci. {\bf 5}, 17--61.

\item S.N. Evangelou, 1983,
   {\em Quantum percolation and the Anderson transition in dilute systems},
   Phys. Rev. B {\bf 27}, 1397--1400.

\item I.J. Farkas, I. Der\'enyi, A.L. Barab\'asi and T. Vicsek, 2001, 
   {\em Spectra of ``real-world'' graphs: beyond the semicircle law},
   Phys. Rev. E {\bf 64}, 026704.

\item S. Kirkpatrick and T.P. Eggarter, 1972,
   {\em Localized states of a binary alloy},
   Phys. Rev. B {\bf 6}, 3598--3609.

\item D.E. Knuth, 1997,
  {\em The art of computer programming},
  Vol. 1, 3rd edition. 

\item G. Li and F. Ruskey, 1999,
  {\em The advantages of forward thinking in generating rooted and free trees},
  SODA 1999, S939-940.

\item A.D. Mirlin, 2000,
   {\em Statistics of energy levels and eigenfunctions in disordered
        systems},
   Phys. Rep. {\bf 236}, 259--382.

\item J.H. van Lint and R.M. Wilson, 1992,
   {\em A Course in Combinatorics}, Cambridge University Press, New York.

\item R.A. Wright, B. Richmond, A. Odlyzko, and B.D. McKay, 1986,
  {\em Constant time generation of free trees}, 
  SIAM J. Comput., {\bf 15}, 540-548. 

\end{itemize}

\end{document}